\documentclass[english]{article}
\usepackage[T1]{fontenc}
\usepackage[latin1]{inputenc}
\usepackage{babel}
\usepackage{graphics}

\makeatletter

\providecommand{\LyX}{L\kern-.1667em\lower.25em\hbox{Y}\kern-.125emX\@}

 \newcommand{\lyxaddress}[1]{
   \par {\raggedright #1 
   \vspace{1.4em}
   \noindent\par}
 }

\makeatother
\begin{document}

\title{A Micro-Canonical Description of Hadron Production \\
in Proton-Proton Collisions}

\author{F. M. Liu\( ^{1,2,3} \)\footnote{%
Fellow of the Alexander von Humboldt Foundation}%
, K. Werner\( ^{3} \), J. Aichelin\( ^{3} \), M. Bleicher\( ^{2} \),
H. Stöcker\( ^{2} \)}

\maketitle

\lyxaddress{\( ^{1} \)Institute of Particle Physics, Central China Normal University,
Wuhan, China}

\lyxaddress{\( ^{2} \)Institut fuer Theoretische Physik, J. W. Goethe-Universitaet,
Frankfurt am Main, Germany}

\lyxaddress{\( ^{3} \)Laboratoire SUBATECH, Universite de Nantes - IN2P3/CNRS
- Ecole des Mines de Nantes, Nantes, France}

\begin{abstract}
A micro-canonical treatment is used to study particle production in
pp collisions. First this micro-canonical treatment is compared to
some canonical ones. Then proton, antiproton and pion 4\( \pi  \)
multiplicities from proton-proton collisions at various center of
mass energies are used to fix the micro-canonical parameters \( E \)
and \( V \). The dependences of the micro-canonical parameters on
the collision energy are parameterised for the further study of pp
reactions with this micro-canonical treatment. 
\end{abstract}
In recent years, the use of thermal statistical models to describe
particle ratios for heavy ion collisions has received great attention\cite{Fermi:1950jd,Landau,Hagedorn,Siemens,Mekjian,Csernai,Stoecker,Rafelski}.
Thermal models are even used to describe proton-proton and \( \mathrm{e}^{+}\mathrm{e}^{-} \)
collisions. The corresponding volumes for these reactions are very
small (\( V\sim 25\, fm^{3} \)). For small volumes, energy and flavour
conservation become important. Therefore a micro-canonical treatment
should be employed. With the increase of the volume, the micro-canonical
results should converge towards the canonical one.

The following micro-canonical treatment \cite{wer95} is employed.
Following the general philosophy of statistical approaches to hadron
production, we suppose that the result of a high energy collision
can be considered as a distribution of {}``clusters'', {}``droplets'',
or {}``fireballs'', which move relative to each other. Here, we
are only interested in \( 4\pi  \) particle yields and average transverse
momenta, and the distribution of clusters may be identified with one
single {}``equivalent cluster'', being characterised by its volume
\( V \) (the sum of individual proper volumes), its energy \( E \)
(the sum of all the cluster masses), and the net flavour content \( Q=(N_{u}-N_{\bar{u}},N_{d}-N_{\bar{d}},N_{s}-N_{\bar{s}}) \).
The basic assumption is that a cluster, characterised by \( V \),
\( E \), and \( Q \), decays {}``statistically'' according to
phase space. More precisely, the probability of a cluster to hadronize
into a configuration \( K=\{h_{1},\ldots ,h_{n}\} \) of hadrons \( h_{i} \)
is given by the micro-canonical partition function \( \Omega (K) \)
of an ideal, relativistic gas of the \( n \) hadrons \( h_{i} \),\[
\Omega (K)=\frac{V^{n}}{(2\pi \hbar )^{3n}}\, \prod _{i=1}^{n}g_{i}\, \prod _{\alpha \in \mathcal{S}}\, \frac{1}{n_{\alpha }!}\, \int \prod _{i=1}^{n}d^{3}p_{i}\, \delta (E-\Sigma \varepsilon _{i})\, \delta (\Sigma \vec{p}_{i})\, \delta _{Q,\Sigma q_{i}},\]
with \( \varepsilon _{i}=\sqrt{m_{i}^{2}+p_{i}^{2}} \) being the
energy, and \( p_{i} \) being the 3-momentum of particle \( i \).
The term \( \delta _{Q,\Sigma q_{i}} \) ensures flavour conservation;
\( q_{i} \) is the flavour vector of hadron \( i \). The symbol
\( \mathcal{S} \) represents the set of hadron species considered:
we take \( \mathcal{S} \) to contain the pseudoscalar and vector
mesons \( (\pi ,K,\eta ,\eta ',\rho ,K^{*},\omega ,\phi ) \) and
the lowest spin-\( \frac{1}{2} \) and spin-\( \frac{3}{2} \) baryons
\( (N,\, \Lambda ,\, \Sigma ,\, \Delta ,\, \Sigma ^{*},\, \Xi ^{*},\, \Omega ) \)
and the corresponding antibaryons. \( n_{\alpha } \) is the number
of hadron species \( \alpha  \), and \( g_{i} \) is the degeneracy
of particle \( i \).

For the present investigation, we limit the number of hadrons in \( \mathcal{S} \)
to 54. Strange particles are produced according to phase space, i.e.
without applying any artificial suppression factor. The results are
compared to that of the canonical calculations using the approach
of Becattini et al. \cite{bec1,becee,becpp}, but without strangeness
suppression and the number of hadrons which can be produced is limited
to the same 54 species allowed in our micro-canonical treatment. In
principle it is possible to include more hadrons in the micro-canonical
ensemble. This makes a detailed comparison more difficult, because
the less known decay channels of the additional hadrons have to agree. 

\begin{figure}
{\centering \resizebox*{0.9\textwidth}{!}{\includegraphics{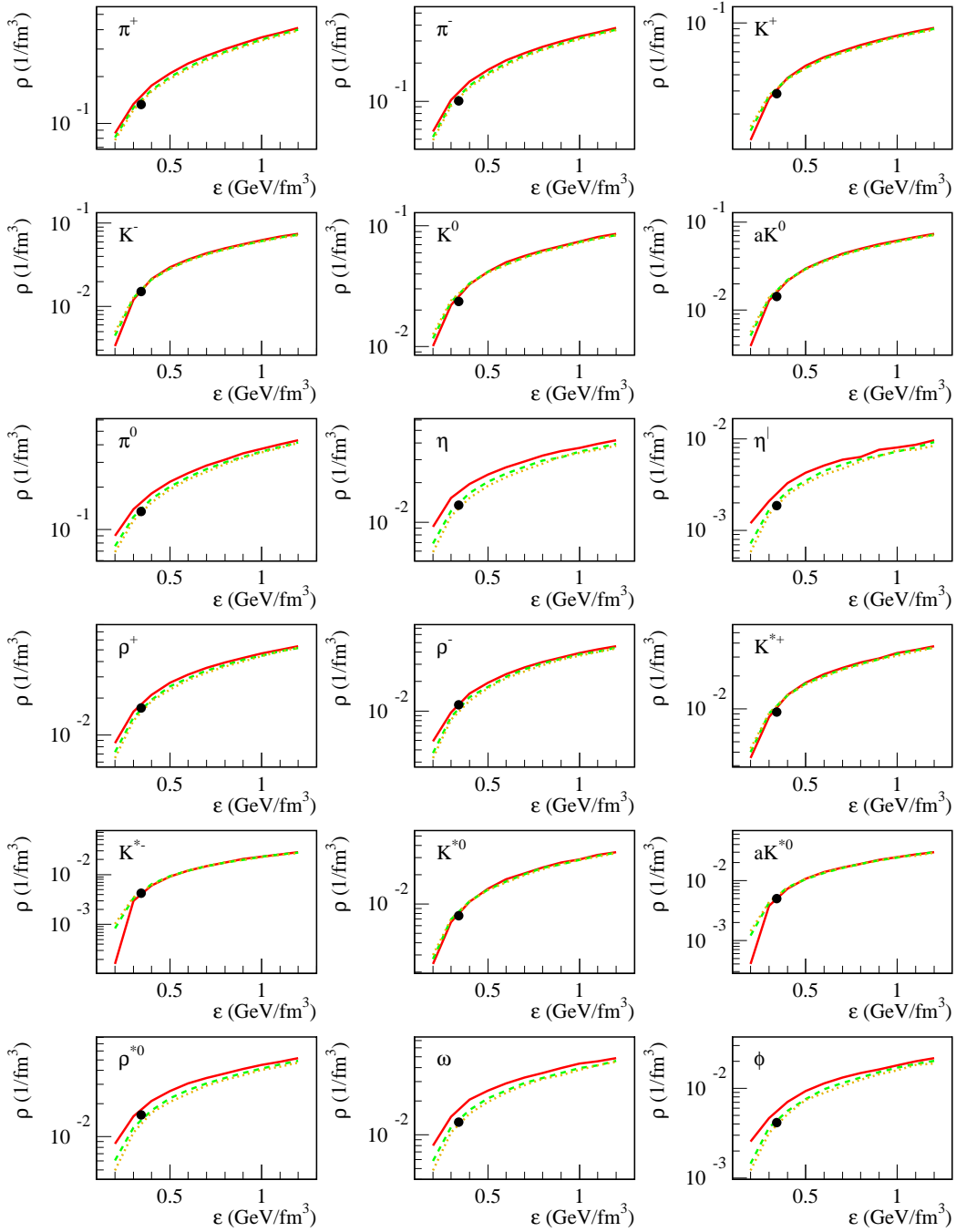}} \par}

\caption{\label{fig1a}Density of particles as a function of the energy density
of a cluster for three different volumes: 12.5 \protect\( fm^{3}\protect \)
(dashed line) and 50 \protect\( fm^{3}\protect \) and a total charge
of 2 using a micro-canonical phase space calculation. The dots present
the result of a canonical calculation\cite{bec1,becee,becpp} provided
by Becattini. In both cases the number of hadrons is limited to 54
and strange particles are not suppressed. Taken from \cite{fuming03a}.}
\end{figure}

\begin{figure}
{\centering \resizebox*{0.9\textwidth}{!}{\includegraphics{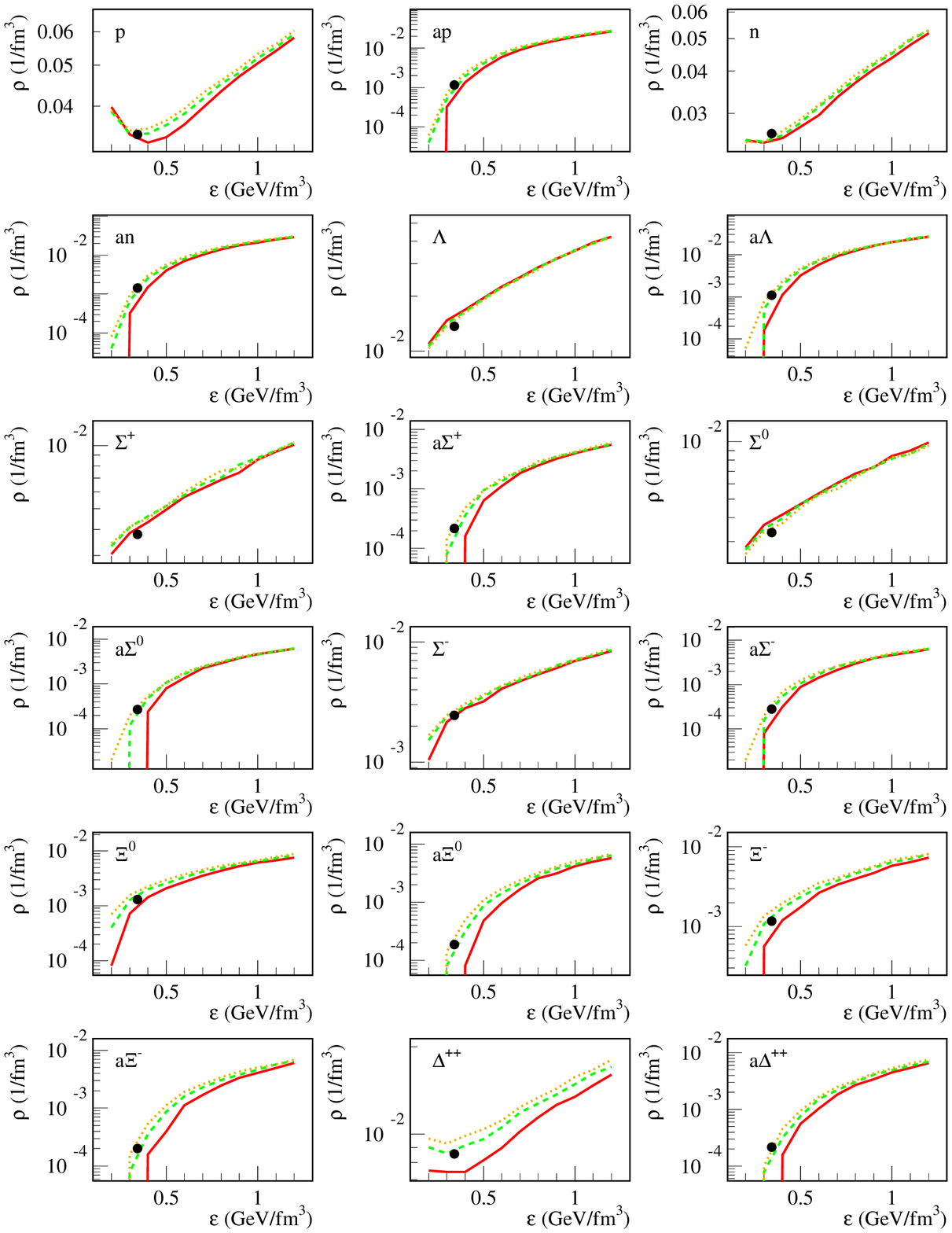}} \par}

\caption{\label{fig1b}Same as Fig. \ref{fig1a}, but for additional hadrons}
\end{figure}

\begin{figure}
{\centering \resizebox*{0.9\textwidth}{!}{\includegraphics{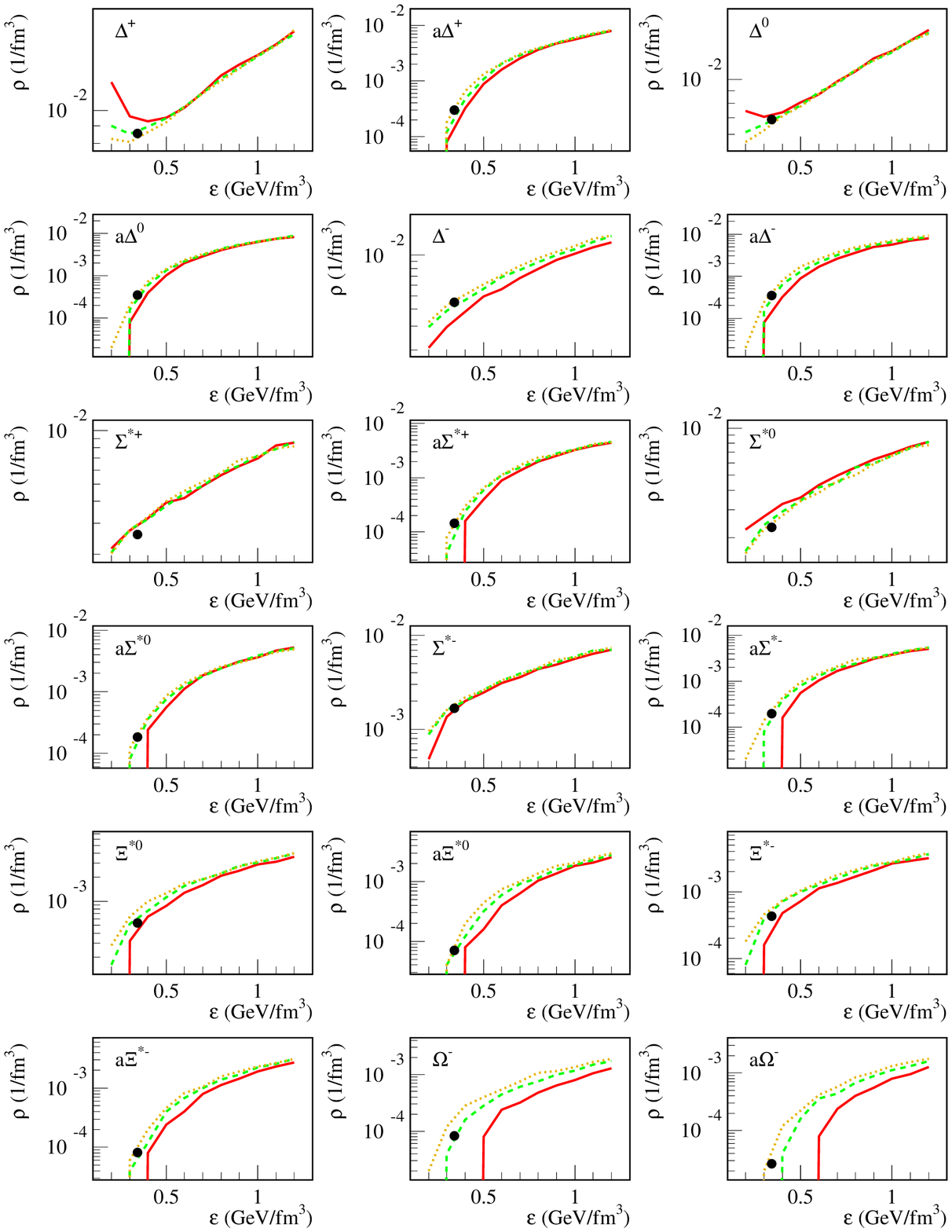}} \par}

\caption{\label{fig1c}Same as Fig \ref{fig1a}, but for additional hadrons}
\end{figure}

For micro-canonical treatment, there exists a critical volume \( V_{c} \),
above which the physics results, i.e. particle yield per volume and
average transverse momentum, become independent of volume and coincide
to the canonical ones. In the following we check the volume effect
in particle yields, compare the micro-canonical results to canonical
one and try to find the critical volume \( V_{c} \). To do this calculation,
we keep the same baryon density as input, i.e. the baryon number is
two (a proton-proton collision) for the volume \( 25\, \mathrm{fm}^{3} \)
which is usually used by canonical calculations (dashed lines), baryon
number one for volume \( 12.5\, \mathrm{fm}^{3} \) (solid lines)
and baryon number four for volume \( 50\, \mathrm{fm}^{3} \) (dotted
lines), c.f. Figs \ref{fig1a}, \ref{fig1b}, \ref{fig1c}. The results
of a canonical calculation with the parameters fitted to describe
the particle yields observed in pp collisions at 27.4 GeV \cite{bec1,becee,becpp}
are marked as dot points. The parameters used in this canonical calculations
are  \( V=25.5\, \mathrm{fm}^{3} \) and \( T=162\, \mathrm{MeV} \).
In contradistinction to the calculation with this parameter which
is presented in ref. \cite{becpp} here the strange particles are
not suppressed by a \( \gamma _{s} \) factor and only the 54 hadron
species mentioned above are produced. Therefore this canonical calculation
can directly be compared with our micro-canonical approach. The average
energy of the clusters obtained in this calculation is 8.74 GeV, resulting
in an average energy density of \( \varepsilon =0.342\, \mathrm{GeV}/\mathrm{fm}^{3} \).
More detailed explanation of Figs \ref{fig1a}, \ref{fig1b},\ref{fig1c}
can be found in \cite{fuming03a}.

Figs \ref{fig1a}, \ref{fig1b}, \ref{fig1c} tell us, the lighter
is the partcile, the smaller is the critical volume \( V_{c} \).
90\% newly-produced hadrons from the collisions are pions while the
critical volume \( V_{c} \) is not bigger than \( 12.5\, \mathrm{fm}^{3} \).
However, for the heavy particles, i.e. \( \Delta ^{++} \) and \( \Omega  \),
the critical volume \( V_{c} \) is bigger than \( 50\, \mathrm{fm}^{3} \).
The canonical results agree well with the micro-canonical one at the
same volume, \( 25\, \mathrm{fm}^{3} \). Therefore we raise the question
if it is safe to study the heavy particles from small reaction systems
with canonical models. 

Canonical calculations get volume \( 25\, \mathrm{fm}^{3} \) from
fitting particle yields. Here we also fit the \( 4\pi  \) particle
yields from pp collisions to fix the micro-canonical parameters \( E \)
and \( V. \) The micro-canonical model does not have any strangeness
suppression factor. It fails to describe non-strange and strange hadrons
at the same time. So we only consider non-strange hadron production
here. To fit the data, we minimize 

\[
\chi ^{2}=\frac{1}{\alpha }\sum ^{\alpha }_{j=1}\frac{[n_{\mathrm{exp},j}(\sqrt{s})-n_{\mathrm{th},j}(E,V)]^{2}}{\sigma _{j}^{2}}\]
to determine the parameter \emph{E} and \emph{V} for at each collision
energy \( \sqrt{s} \), where \( n_{\mathrm{exp},j} \), \( n_{\mathrm{th},j} \)
and \( \sigma _{j} \) are respectively the experimental multiplicity,
multiplicity from the micro-canonical treatment and experimental variance
of particle \( j \). Both the experimental and theoretical multiplicities
here are after decays. We take the parameterization\cite{antinucci}
of energy dependence of \( \pi ^{+} \), \( \pi ^{-} \), proton and
antiproton instead of the data to fix the micro-canonical parameters
\( E \) and \( V \) for a convenient reason.
\begin{figure}
{\centering \resizebox*{0.8\textwidth}{!}{\includegraphics{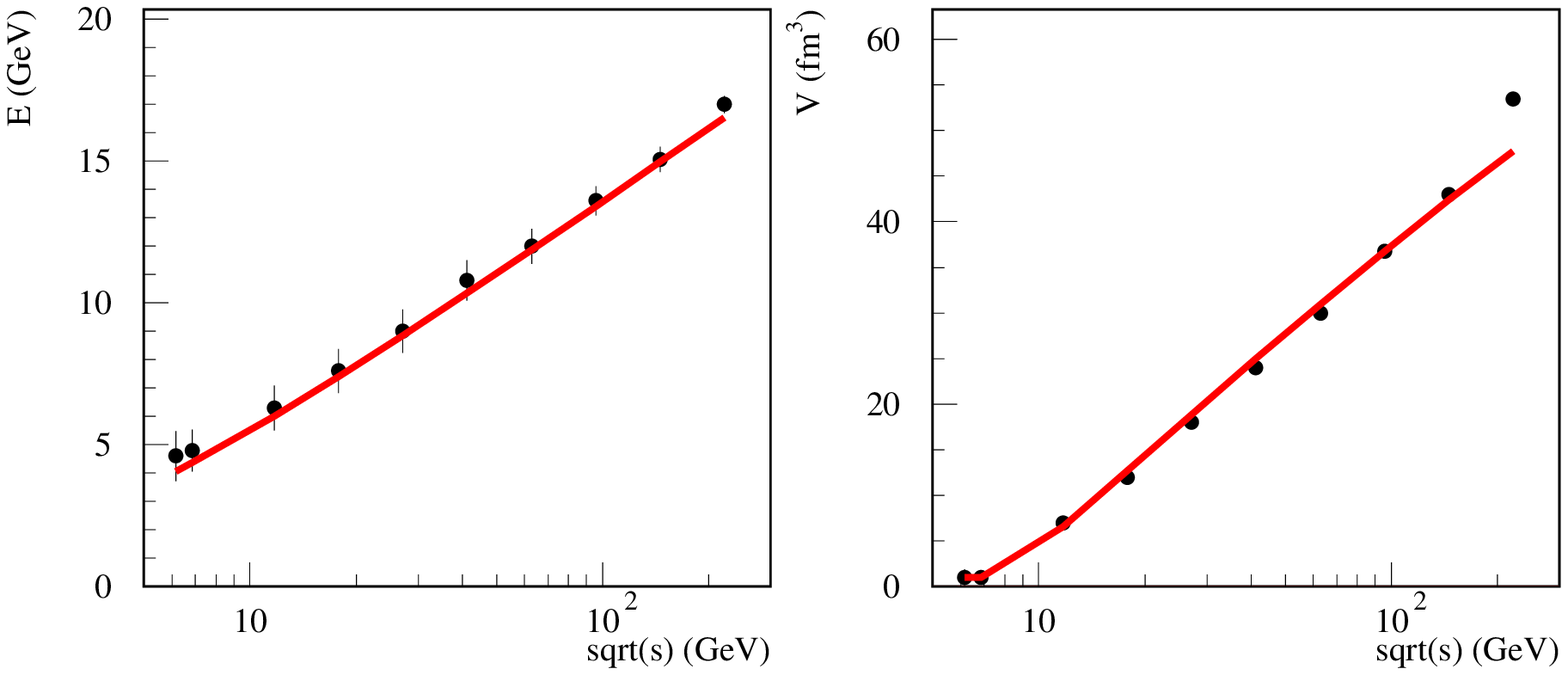}} \par}

\caption{\label{fitdepp}Micro-canonical parameters \protect\( E\protect \)
(left) and \protect\( V\protect \) (right) dependence on the collision
energy \protect\( \sqrt{s}\protect \). The points are from fitting.
The solid lines are the parameterized function described in Eq. \ref{e}
and \ref{v}.}
\end{figure}
The minimum \( \chi ^{2} \) method produces the micro-canonical parameters
\( E \) and \( V \) dependence on collision energy \( \sqrt{s} \)
as shown in Fig. \ref{fitdepp}.  \( E \) and \( V \) increase monotonously
with \( \sqrt{s} \) . We parameterize the micro-canonical parameters
\( E \) (GeV) and \( V \) (\( \mathrm{fm}^{3} \)) dependence on
the collision energy \( \sqrt{s} \) (GeV) for some further study
of pp cillisions with this canonical treatment: \begin{eqnarray}
E & = & -3.8+3.76\mathrm{ln}\sqrt{s}+6.4/\sqrt{s}\label{e} \\
V & = & -30.04+14.93\mathrm{ln}\sqrt{s}-0.013\sqrt{s}\label{v} 
\end{eqnarray}
Note that \( V_{\mathrm{min}}=1\, \mathrm{fm}^{3} \) has been used
for Eq. \ref{v} at very low energies. The Eq. \ref{e} itself is
the same as charged hadron excitation function in pp collisions\cite{antinucci}.
This agrees well with the results in heavy ion collsions\cite{Cleymans}
where the freeze-out energy per particle \( \left\langle E\right\rangle /\left\langle N\right\rangle =1\, \mathrm{GeV} \),
taking into account that the increase of the multiplicity from decay
has roughly the same fraction of neutral particles in the total hadron
multiplicity.

\noindent \textbf{Acknowledgment}

We would like to thank F. Becattini for many discussions and for providing
us with the calculation with a restricted set of hadrons and without
strangeness suppression.


\begin{thebibliography}{10}
\bibitem{Fermi:1950jd}E. Fermi, Prog. Theor. Phys. \textbf{5}, 570 (1950); Phys.Rev. 81
(1951) 683.
\bibitem{Landau}L. D. Landau, Lzv. Akd. Nauk SSSR 17 (1953) 51; Collected papers of
L. D. Landau, ed. D. Ter Haar, Gordon and Breach, New York, 1965
\bibitem{Hagedorn}R. Hagedorn, Nucl. Phys. B \textbf{24} (1970) 93. 
\bibitem{Siemens}P. Siemens, J. Kapusta, Phys. Rev. Lett. \textbf{43} (1979) 1486.
\bibitem{Mekjian}A. Z. Mekjian, Nucl. Phys. A\textbf{384} (1982) 492.
\bibitem{Csernai}L. Csernai, J. Kapusta Phys. Rep. \textbf{131} (1986) 223.
\bibitem{Stoecker}H. Stoecker, W. Greiner, Phys. Rep. \textbf{137} (1986) 279.
\bibitem{Rafelski}J. Rafelski and J. Letessier, J. Phys. G: Nucl. Part. Phys. \textbf{28}
(2002) 1819-1832.
\bibitem{wer95}K. Werner and J. Aichelin, Phys. Rev. \textbf{C52} (1995) 1584-1603.
\bibitem{bec1}F. Becattini, G. Passaleva, Eur.Phys.J. \textbf{C23} (2002) 551 
\bibitem{becpp}F. Becattini , U. Heinz Z.Phys. \textbf{C76} (1997) 269-286 
\bibitem{becee}F. Becattini , A. Giovannini, S. Lupia Z.Phys. \textbf{C72} (1996)
491
\bibitem{fuming03a}F.M. Liu, K. Werner, J. Aichelin, hep-ph/0304174, Comparison of micro-canonical
and Canonical Hadronization.
\bibitem{antinucci}Antinucci et al, Letters al Nuovo Cimento, \textbf{V6}, N4, 27 Gennaio1973.
\bibitem{Cleymans}J. Cleymans and K. Redlich, Phys.Rev.Lett. \textbf{81}:5284-5286,1998.\end{thebibliography}
\end{document}